\documentclass[submission,copyright,creativecommons]{eptcs}
\providecommand{\event}{GASCom 2026} 

\usepackage{iftex}

\ifpdf
  \usepackage{underscore}         
  \usepackage[T1]{fontenc}        
\else
  \usepackage{breakurl}           
\fi

\usepackage{graphicx} 
\usepackage[english]{babel}
\usepackage{amsthm}
\usepackage[normalem]{ulem}
\usepackage{amssymb}
\usepackage[toc,page]{appendix}
\usepackage{algorithm}
\usepackage{amsmath}
\usepackage[noend]{algpseudocode}
\usepackage{enumitem}
\usepackage{xcolor}
\usepackage{multicol}
\makeatletter
\def\BState{\State\hskip-\ALG@thistlm}
\usepackage{hyperref}
\def\Av{\mathrm{Av}}
\def\rank{\mathrm{rank}}
\def\unrank{\mathrm{unrank}}
\makeatother
\newtheorem{theorem}{Theorem}[section]

\newtheorem{proposition}[theorem]{Proposition}

\theoremstyle{definition}
\newtheorem{definition}[theorem]{Definition} 
 
\newtheorem{rk}[theorem]{Remark}
\newtheorem{fact}[theorem]{Fact}

\title{(Un)ranking Permutation Classes}
\author{Nathanaël Hassler \qquad\qquad Vincent Vajnovszki
\institute{Laboratoire d'Informatique de Bourgogne\\ 
Université Bourgogne Europe\\
Dijon, France}
\email{\quad nathanael.hassler@ens-rennes.fr \quad\qquad vincent.vajnovszki@u-bourgogne.fr}
}
\def\titlerunning{(Un)ranking Permutation Classes}
\def\authorrunning{N. Hassler \& V. Vajnovszki}
\begin{document}
\maketitle

\begin{abstract}
Permutations avoiding a pattern of length three are enumerated by the Catalan numbers. In this work, we present methods for ranking and unranking such permutations in lexicographic or colexicographic order.
\end{abstract}

\section{Introduction}
For an ordered set of combinatorial objects we can ask which position in the ordered list of objects a given object has, or conversely determine the objet in a given position. Formally for a total ordered set $C$ we have:

\begin{definition}
The {\em ranking} over $C$ is the function
$$
\rank_{\,C} : C \rightarrow \{0,1,\dots,|C|-1\}
$$
that maps $x$ to the number of elements of $C$ that are less than $x$.
The {\em unranking} is the inverse function
$$
\unrank_{\,C} : \{0,1,\dots,|C|-1\} \rightarrow C
$$
that maps $p$ to the object $x$ such that $\rank_{\,C}(x)=p$.
\end{definition}

Here, we restrict ourselves to sets $C$ of restricted permutations of the same length in one-line representation, ordered lexicographically or colexicographically. Whenever the set $C$ is clear from the context, the subscripts in the functions $\rank$ and $\unrank$ will be omitted.

Lehmer \cite{Lehmer} may have been the first to suggest studying combinatorial objects by imposing an order on them and then developing corresponding ranking and unranking algorithms. Ruskey in his book \cite{Ruskey} considers such algorithms for unrestricted permutations in both (co)lexicographic order and Steinhaus–Johnson–Trotter Gray code order. Among recent contributions in the field, we can cite \cite{CurielGenitrini,GenitriniPepin}, and more particularly \cite{Kagey} for ranking and unranking algorithms for restricted permutations (derangements and ménage permutations).

Here, we are interested in ranking and unranking algorithms for simple yet non-trivial combinatorial classes, namely permutations avoiding a pattern of length three.

Aside from their intrinsic theoretical interest, such methods also have practical value.
Indeed, given an unranking method, the ordered list can be computed as $$\unrank(0),\unrank(1),\unrank(2),\dots$$
and in the context of random sampling, $\unrank(p)$ yields a random object as long as $p$ is a random integer less than the cardinality of the combinatorial class. In addition, having a ranking method, the successor of a given object $x$ can be computed as $\unrank(\rank(x)+1)$. Even though these techniques are not efficient in general, they can provide hints for solutions to the underlying problems and enhance understanding of the concerned objects.

A permutation of the set $[n] = \{1, 2, \dots, n\}$ is a bijection from $[n]$ onto itself, and 
$\mathfrak{S}_n$ denotes the set of such permutations. We will represent permutations by their one-line notation, that is, the permutation $\pi \in \mathfrak{S}_n$ by the word $\pi(1)\pi(2)\cdots \pi(n) \in [n]^n$.

Let $k$ and $n$ be integers, $0<k\leq n$, and $\sigma=\sigma(1)\sigma(2)\cdots \sigma(k)\in \mathfrak{S}_k$ and $\pi=\pi(1)\pi(2)\cdots \pi(n)\in \mathfrak{S}_n$ be two permutations. One says that $\pi$ contains $\sigma$ if $\pi$ contains a (not necessarily contiguous) subsequence $\pi(i_1)\pi(i_2)\cdots\pi(i_k)$, $i_1 < i_2 <\cdots < i_k$, order isomorphic to $\sigma$; otherwise one says that $\pi$ avoids $\sigma$, or $\pi$ is $\sigma$-avoiding. In this context $\sigma$ is called a (classical) {\em pattern}, and the set of length-$n$ permutations avoiding the pattern $\sigma$ is denoted by $\Av_n(\sigma)$, and $\Av(\sigma)=\cup_{n\geq 0}\Av_n(\sigma)$.

The lexicographic order on $[n]^n$ induces the lexicographic order on permutations in $\mathfrak{S}_n$ in their one-line representation, and it generalizes to words over any ordered alphabet. For two same length words $v=v_1v_2\dots v_n$ and $w=w_1w_2\dots w_n$ one says that $v$ is less than $w$ in {\em colexicographic order} if 
$v_nv_{n-1}\dots v_1$ is less than $w_nw_{n-1}\dots w_1$ in lexicographic order.

In the following we need the next easy to check result.

\begin{rk}
If $C$ is a set of length-$n$ words over an ordered alphabet, then the rank of $w=w_1w_2\dots w_n\in C$, in lexicographic order, is
\begin{equation}
\rank(w)=
\sum_{i=1}^{n}|E_i|,
\end{equation}
where for each $i$, $E_i$ is the set of words $u_1u_2\dots u_n\in C$ with 
$u_j=w_j$ for $j<i$ and $u_i<w_i$.
\end{rk}

The number of permutations in $\mathfrak{S}_n$ avoiding a pattern of length three is given by the Catalan number $c_n=\frac{1}{n+1}{{2n}\choose{n}}$, see  \cite{Knuth}, which is sequence \href{https://oeis.org/A000108}{A000108} in~\cite{OEIS}.
Often it is convenient to represent 
pattern avoiding 
permutations $\pi\in \mathfrak{S}_n$ by an $n\times n$ array with a dot in each one of the squares $(i,\pi(i))$. See Figures \ref{Fig3perms} (b) and (c) for two examples.

\section{Avoiding the $231$ pattern}

The {\em reduction} of a word $\alpha$ over a finite alphabet $A \subset \mathbb{N}$ is obtained by replacing each occurrence of the smallest symbol by 1, each occurrence of the second smallest symbol by 2, and so on. The {\em direct sum} of two permutations $\sigma\in \mathfrak{S}_k$ and $\tau\in \mathfrak{S}_\ell$, denoted by $\sigma\oplus\tau$, is the permutation $\pi\in \mathfrak{S}_{k+\ell}$ where $\pi(1)\pi(2)\cdots \pi(k)=\sigma$ and 
$\pi(k+1)\pi(k+2)\cdots \pi(k+\ell)$ is 
a permutation of $\{k+1,k+2,\dots, k+\ell\}$ that reduces to $\tau$. Similarly, the {\em skew sum} of $\sigma$ and $\tau$, denoted by $\sigma\ominus\tau$, is the permutation $\pi\in \mathfrak{S}_{k+\ell}$ where $\pi(1)\pi(2)\cdots \pi(k)$ is a permutation of $\{\ell+1,\ell+2,\dots, \ell+k\}$ that reduces to $\sigma$ and
$\pi(k+1)\pi(k+2)\cdots \pi(k+\ell)=\tau$. 
With this notation, if $\pi = 1 \ominus \sigma$, we have $\pi(1) = k+1$.

The next easy to understand fact is folklore.

\begin{fact}
A permutation avoids $231$ if and only if it has the form $(1 \ominus \sigma)\oplus \tau$, 
where $\sigma$ and $\tau$ are (possibly empty) $231$-avoiding permutations.
\end{fact}

\begin{figure}
\begin{center}
\begin{tabular}{ c c c }
\setlength{\unitlength}{0.04cm}
\begin{picture}(100,100)
\linethickness{0.01mm}
\put(0,0){\vector(0,1){85}}
\multiput(0,0)(70,0){2}{\line(0,1){70}} 
\put(0,0){\vector(1,0){85}}
\multiput(0,0)(0,70){2}{\line(1,0){70}}
\put(10,0){\line(0,1){20.4}}
\put(10,20){\line(1,0){20.4}}
\put(30,0){\line(0,1){20.4}}
\put(10,0){\line(1,0){20.4}}

\put(30,30){\line(0,1){40.4}}
\put(30,70){\line(1,0){40.4}}
\put(30,30){\line(1,0){40.4}}
\put(70,30){\line(0,1){40.4}}

\put(5,25){\circle*{4}}

\put(-18,35){\scriptsize{$\pi(i)$}}
\put(35,-10){\scriptsize{$i$}}
\put(17,8){\scriptsize{$\sigma$}}
\put(47,44){\scriptsize{$\tau$}}

\end{picture}
& 
\setlength{\unitlength}{0.04cm}
\begin{picture}(100,100)
\linethickness{0.01mm}
\put(0,0){\vector(0,1){85}}
\multiput(0,0)(10,0){8}{\line(0,1){70}} 
\put(0,0){\vector(1,0){85}}
\multiput(0,0)(0,10){8}{\line(1,0){70}}
\put(5,15){\circle*{4}}
\put(15,5){\circle*{4}}
\put(25,65){\circle*{4}}
\put(35,25){\circle*{4}}
\put(45,45){\circle*{4}}
\put(55,35){\circle*{4}}
\put(65,55){\circle*{4}}
\put(-18,35){\scriptsize{$\pi(i)$}}
\put(35,-10){\scriptsize{$i$}}
\end{picture}

&
\setlength{\unitlength}{0.04cm}
\begin{picture}(100,100)
\linethickness{0.01mm}
\put(0,0){\vector(0,1){85}}
\multiput(0,0)(10,0){8}{\line(0,1){70}} 
\put(0,0){\vector(1,0){85}}
\multiput(0,0)(0,10){8}{\line(1,0){70}}
\linethickness{0.35mm}
\put(0,0){\line(0,1){20.4}}
\put(0,20){\line(1,0){20}}
\put(20,19.7){\line(0,1){30.7}}
\put(20,50){\line(1,0){20.3}}
\put(40,50){\line(0,1){20.5}}
\put(40,70){\line(1,0){30}}
\put(5,15){\circle*{4}}
\put(25,45){\circle*{4}}
\put(45,65){\circle*{4}}
\put(15,5){\circle*{4}}
\put(35,25){\circle*{4}}
\put(55,35){\circle*{4}}
\put(65,55){\circle*{4}}
\put(-18,35){\scriptsize{$\pi(i)$}}
\put(35,-10){\scriptsize{$i$}}
\end{picture}\\  
\\
 (a) & (b)  & (c)   
\end{tabular}
\end{center}
\caption{The arrays  representation of: (a) the permutation $\pi=(1\ominus \sigma)\oplus \tau$; (b) the 231-avoiding  permutation 2173546; (c) the 321-avoiding permutation 2153746 corresponding to the Dyck word $bbaabbbaabbaaa$ which codes the over-diagonal path in bold.
\label{Fig3perms}}
\end{figure}
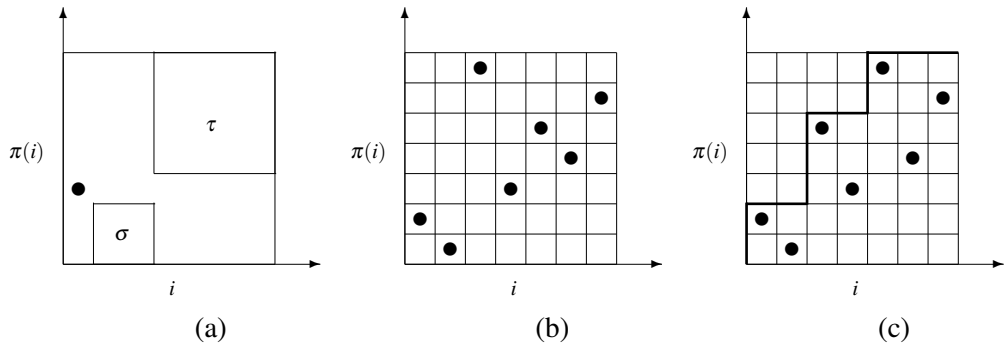

See Figure \ref{Fig3perms} (a) for the general shape of a 231-avoiding permutation, and (b) for the 231-avoiding permutation $2173546=
(1\ominus 1)\oplus (1\ominus (1\oplus ((1\ominus 1)\oplus 1)))$.

\begin{proposition}
\label{propRank231}
Let $\pi=\pi(1)\pi(2)\cdots \pi(n)=(1\ominus \sigma)\oplus \tau$ be a length-$n$ 231-avoiding permutation, with $\sigma\in \Av_k(231)$ and $\tau\in \Av_\ell(231)$ (and thus, $n=k+\ell+1$). Then the rank of $\pi$, in lexicographical order, is given recursively by
$$
\rank_{\Av_n(231)}(\pi)=\sum_{i=1}^{\pi(1)-1} c_i\cdot c_{n-i-1} + \rank_{\Av_k(231)}(\sigma)\cdot 
c_\ell + \rank_{\Av_\ell(231)}(\tau).$$
\end{proposition}


\begin{proposition}
\label{propUnrank231}
Let $n,p$ be integers, $0\leq p<c_n$. The permutation $\unrank_{\Av_n(231)}(p)$ is obtained recursively as follows.
Determine the smallest $k$ such that 
$$
p< \sum_{i=0}^kc_ic_{n-i-1}
$$
and denote
\begin{itemize}
\item[] $s:=\sum_{i=0}^{k-1} c_ic_{n-i-1}$,
\item[] $u:=\lfloor \frac{p-s}{c_{n-k}}\rfloor$, and
\item[] $v:=p-s-u\cdot c_{n-k}$.
\end{itemize}
Then $\unrank_{\Av_n(231)}(p)=(1\ominus \sigma)\oplus\tau$
where 
\begin{itemize}
\item[] 
$\sigma=\unrank_{\Av_{k-1}(231)}(u)$, and 
\item[]
$\tau=\unrank_{\Av_{n-k}(231)}(v)$.
\end{itemize}
\end{proposition}

In fact, the value $k$ defined in the previous proposition is the first entry of $\unrank_{\Av_n(231)}(p)$.

\section{Avoiding the $321$ pattern}

A \emph{left-to-right maximum position} (or \emph{lrmp} for short) in a permutation $\pi$ is an index $i$ such that, for all $j < i$, we have $\pi(j) < \pi(i)$. The value $\pi(i)$ is then called a \emph{left-to-right maximum value} (or \emph{lrmv} for short). An \emph{excedance} is an index $i$ such that $\pi(i) > i$.
It is known (see, e.g., \cite{Reifegerste}) that a permutation is $321$-avoiding if and only if both the subsequence of its left-to-right maximum values and the subsequence of its remaining elements are increasing.
Moreover, such a permutation is uniquely determined by the set $\{(i,\pi(i))\}$, where $i$ ranges over its left-to-right maximum positions. Even though we will not use it later on, it is worth mentioning that, for a $321$-avoiding permutation, the left-to-right maxima positions coincide with the excedances of the permutation.

Another combinatorial class closely related to length-3 pattern-avoiding permutations is that of \emph{Dyck words}. Let $D_n$ denote the set of Dyck words of semilength $n$, that is, words 
$w \in \{a,b\}^{2n}$ containing the same number 
of occurrences of $a$ and $b$, and such that every prefix of $w$ 
contains at least as many $b$’s as $a$’s. 
(We adopt this unusual notation for convenience; traditionally, the roles of $a$ and $b$ are interchanged.) Dyck words in $D_n$ are counted by the Catalan number $c_n$.

The bijection $\psi:\Av_n(321)\to D_n$ appeared originally in \cite[p. 89]{Knuth}
in a slightly different form.
Consider the array representation corresponding to $\pi\in \Av_n(321)$ and the path with {\em east} and {\em north} steps along
the edges of the array 
that goes from the lower-left corner to the upper-right corner of the array, leaving all the dots to the right and remaining always as close to the
main diagonal as possible. Let $P$ be the resulting path. Then $\psi(\pi)$ can be obtained from $P$ just
by reading its steps, a north step corresponding to the letter $b$ and an east step to the letter $a$. See Figure \ref{Fig3perms} (c) for an example.

The next proposition says that $\psi$ preserves the lexicographic order.

\begin{proposition}
If $\pi,\tau\in \Av_n(321)$ with $\pi<\tau$ in lexicographic order, then $\psi(\pi)<\psi(\tau)$ in lexicographic order as well.
\end{proposition}


{\em Ballot sequences} are prefixes of Dyck words. We denote by $B(i,j)$ the set of ballot sequences with $i$ occurrences of $b$ and $j$ occurrences of $a$. Similarly, we denote by $\bar{B}(i,j)$ the set of {\em suffixes of Dyck words}.

Clearly, $B(i,j)$ is empty if $i<j$, and  there is a unique word in $B(i,0)$, namely $b^i$.
Moreover, if we ask to obtain a ballot sequence $w\in B(i,j)$ by appending a single letter to a shorter one, then
\begin{itemize}
\item
if $i=j$, this can be obtained in a unique way: $w=ua$ for some $u\in B(i,j-1)$, and
\item if $0<j<i$, this can be obtained in two different ways, namely $w=ua$, for some $u\in B(i,j-1)$, and $w=ub$, for some $u\in B(i-1,j)$.
\end{itemize}

The considerations above (see for instance \cite{Kasa} and the references therein) lead us to the following recursive definition of \( t(i, j) \), the cardinality of \( B(i, j) \):
\begin{equation}
t(i,j)=\left\{
\begin{array}{lll}
0 &  \mathrm{if} & j>i,\\
1 & \mathrm{if} & j=0,\\
t(i,j-1) & \mathrm{if} & j=i>0,\\
t(i-1,j)+t(i,j-1) & \mathrm{if} & 0<j<i.
\end{array}
\right.
\label{recurenceT}
\end{equation}

The sequence $\bigl(t(i, j)\bigr)_{0\leq j\leq i}$
is the sequence 
\href{https://oeis.org/A009766}{A009766} in~\cite{OEIS}, which has the closed form:
\begin{equation}
t(i,j)={i+j\choose
i}\frac{i-j+1}{i+1}, \mbox{ for } 0\leq j\leq i,
\end{equation}
and its first values are shown in Table~\ref{DyckArray}.

\begin{proposition}[\cite{Bailey}]
For the double indexed sequence $s(i,j)=\sum_{k=0}^j t(i,k)$ we have 

$$s(i,j)=t(i+1,j).$$
\end{proposition}

\begin{table}
\begin{center}
\begin{tabular}{ c | c c c c c c c c }
{$i$ \textbackslash $j$}
  & 0 & 1 & 2 & 3 & 4 & 5 & 6 & 7\\
\hline
0 & 1  \\ 
1 & 1 & 1 \\  
2 & 1 & 2 & 2\\
3 & 1 & 3 & 5  & 5\\
4 & 1 & 4 & 9  & 14 & 14\\
5 & 1 & 5 & 14 & 28 & 42  & 42\\
6 & 1 & 6 & 20 & 48 & 90  & 132 & 132\\
7 & 1 & 7 & 27 & 75 & 165 & 297 & 429 & 429
\end{tabular}
\end{center}
\caption{The first values of the double indexed sequence $\bigl(t(i, j)\bigr)_{0\leq j\leq i}$.}
\label{DyckArray}
\end{table}

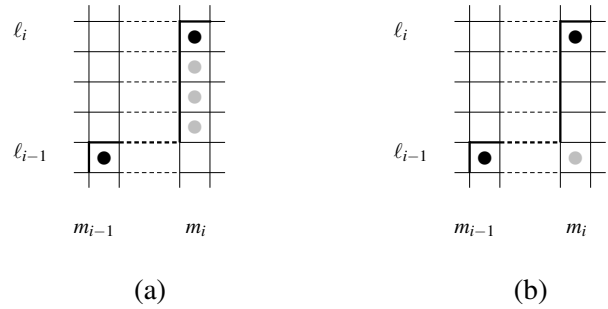
\begin{figure}
\begin{center}
\begin{tabular}{ c c }
\hspace{1cm}
\begin{picture}(100,100)
\setlength{\unitlength}{0.04cm}
\linethickness{0.005mm}

\put(5,30){\line(1,0){15}}
\multiput(20,30)(2.5,0){8}{\line(1,0){1.5}}
\put(40,30){\line(1,0){15}}

\put(5,40){\line(1,0){15}}
\put(40,40){\line(1,0){15}}

\put(5,50){\line(1,0){15}}
\multiput(20,50)(2.5,0){8}{\line(1,0){1.5}}
\put(40,50){\line(1,0){15}}

\put(5,60){\line(1,0){15}}
\multiput(20,60)(2.5,0){8}{\line(1,0){1.5}}
\put(40,60){\line(1,0){15}}

\put(5,70){\line(1,0){15}}
\multiput(20,70)(2.5,0){8}{\line(1,0){1.5}}
\put(40,70){\line(1,0){15}}

\put(5,80){\line(1,0){15}}
\multiput(20,80)(2.5,0){8}{\line(1,0){1.5}}
\put(40,80){\line(1,0){15}}

\put(50,25){\line(0,1){60.4}}
\put(40,25){\line(0,1){60.4}}
\put(20,25){\line(0,1){60.4}}
\put(10,25){\line(0,1){60.4}}

\linethickness{0.3mm}

\put(10,30){\line(0,1){10.4}}
\put(10,40){\line(1,0){10}}
\multiput(20,40)(2.5,0){8}{\line(1,0){1.5}}
\put(40,40){\line(0,1){40.4}}
\put(40,80){\line(1,0){10.4}}
\put(15,35){\circle*{4}}
\put(45,75){\circle*{4}}
\put(45,65){\textcolor{lightgray}{\circle*{4}}}
\put(45,55){\textcolor{lightgray}{\circle*{4}}}
\put(45,45){\textcolor{lightgray}{\circle*{4}}}
\put(-15,75){\scriptsize{$\ell_i$}}
\put(-15,35){\scriptsize{$\ell_{i-1}$}}
\put(5,10){\scriptsize{$m_{i-1}$}}
\put(42,10){\scriptsize{$m_i$}}
\end{picture}
& 
\hspace{1cm}
\begin{picture}(100,100)
\setlength{\unitlength}{0.04cm}
\linethickness{0.005mm}

\put(5,30){\line(1,0){15}}
\multiput(20,30)(2.5,0){8}{\line(1,0){1.5}}
\put(40,30){\line(1,0){15}}

\put(5,40){\line(1,0){15}}
\put(40,40){\line(1,0){15}}

\put(5,50){\line(1,0){15}}
\multiput(20,50)(2.5,0){8}{\line(1,0){1.5}}
\put(40,50){\line(1,0){15}}

\put(5,60){\line(1,0){15}}
\multiput(20,60)(2.5,0){8}{\line(1,0){1.5}}
\put(40,60){\line(1,0){15}}

\put(5,70){\line(1,0){15}}
\multiput(20,70)(2.5,0){8}{\line(1,0){1.5}}
\put(40,70){\line(1,0){15}}

\put(5,80){\line(1,0){15}}
\multiput(20,80)(2.5,0){8}{\line(1,0){1.5}}
\put(40,80){\line(1,0){15}}

\put(50,25){\line(0,1){60.4}}
\put(40,25){\line(0,1){60.4}}
\put(20,25){\line(0,1){60.4}}
\put(10,25){\line(0,1){60.4}}

\linethickness{0.3mm}

\put(10,30){\line(0,1){10.4}}
\put(10,40){\line(1,0){10}}
\multiput(20,40)(2.5,0){8}{\line(1,0){1.5}}
\put(40,40){\line(0,1){40.4}}
\put(40,80){\line(1,0){10.4}}
\put(15,35){\circle*{4}}
\put(45,75){\circle*{4}}
\put(45,35){\textcolor{lightgray}{\circle*{4}}}
\put(-15,75){\scriptsize{$\ell_i$}}
\put(-15,35){\scriptsize{$\ell_{i-1}$}}
\put(5,10){\scriptsize{$m_{i-1}$}}
\put(42,10){\scriptsize{$m_i$}}
\end{picture}
\\
(a) & (b)
\end{tabular}
\end{center}
\caption{\label{twoArrays}
The black dots correspond to two consecutive {\em lrmp}'s, $m_{i-1}$ and $m_i$, of a $321$-avoiding permutation $\pi$, together with their {\em lrmv}'s $\ell_{i-1}=\pi(m_{i-1})$ and $\ell_i=\pi(m_i)$. Let $\sigma$ be another $321$-avoiding permutation of the same length that is smaller than $\pi$ in lexicographic order and has the same first $(i-1)$ {\em lrmp}'s $m_1, m_2, \dots, m_{i-1}$, and the same first $(i-1)$ {\em lrmv}'s, that is 
$\sigma(m_j)=\pi(m_j)$ for $j<i$.
(a) The gray dots indicate the possible values for the $i$th {\em lrmv} of $\sigma$ if $m_i$ is still an {\em lrmp} of $\sigma$, with $\sigma(m_i) < \ell_i$.
(b) The gray dot indicates the value of $\sigma(m_i)$ if $m_i$ is no longer an {\em lrmp} of $\sigma$ (and thus $\ell_{i-1} \geq m_i$).
}
\end{figure}

Representing a $321$-avoiding permutation by both the sequence of its left-to-right maxima and the corresponding Dyck word (see
Figure \ref{twoArrays}), we have:
\begin{proposition} 
\label{propAv321}
Let $\pi$ be a length-$n$ $321$-avoiding permutation, and let $(m_i,\ell_i)_{i=1}^k$ be the sequence where the $m_i$’s are the \emph{lrmp}’s of $\pi$, and $\ell_i=\pi(m_i)$ are the \emph{lrmv}’s.
Then, 

$$\rank_{\Av_n({321})} (\pi)=\sum_{i=1}^k \left( s(n-m_i+1,n-\max\{\ell_{i-1},m_i\})-s(n-m_i+1,n-\ell_i+1)\right),$$
where for convenience $\ell_0=0$. 
\end{proposition}

\begin{algorithm}
\caption{\label{UnrankDyck}
Procedure for computing the Dyck word of semilength $n$ and rank $p$. 
Here $\overline{t}(i,j)=\mathrm{card}(\overline{B}(i,j))$, so $\overline{t}(i,j)=t(j,i)$.}
\begin{minipage}[c]{.46\linewidth}
\begin{tabbing}\hspace{0.5cm}\=\hspace{0.5cm}\= \hspace{0.5cm}\= 
\hspace{0.5cm}\= 
\hspace{0.5cm}\=\\
{\bf procedure} {\sc UnrankDyck}($n$,\,$p$: integer)\\
\> {\bf global} $d$: array\\
\> $i\leftarrow n$, $j\leftarrow n$\\
\> {\bf for} $k$ {\bf from} $1$ {\bf to} $2n$ {\bf do}\\
\>\> {\bf if}  $i=j$ {\bf or} $p\geq \overline{t}(i-1,j)$  {\bf then}\\
\>\>\>$d_k\leftarrow\mathrm b$; $p\leftarrow p-\overline{t}(i-1,j)$; $i\leftarrow i-1$\\
\>\> {\bf else} $d_k\leftarrow \mathrm a$; 
$j\leftarrow j-1$ \\
\>\> {\bf end if} \\
{\bf end procedure}\\
\end{tabbing}
\end{minipage}
\end{algorithm}

Let $\pi$ be a $321$-avoiding permutation of length $n$, and let $w = w_1 w_2 \dots w_{2n}$ be its corresponding Dyck word, that is $w=\Psi(\pi)\in D_n$. Obviously, the number of left-to-right maxima in $\pi$ is equal to the number of occurrences of the factor $ba$ (i.e., number of peaks) in $w$. Moreover, if $w_{j-1} w_j$ is the $i$th occurrence of the factor $ba$, then the $i$th left-to-right maximum $(m_i,\ell_i)$ of $\pi$ is given by
$
m_i = |w_1 w_2 \dots w_j|_a$ and $\ell_i = |w_1 w_2 \dots w_j|_b.
$

\begin{proposition}
\label{propUnrank321}
Let $n,p$ be integers with $0 \le p < c_n$. The permutation $\unrank_{\Av_n(321)}(p)$ is obtained as follows:

\begin{itemize}
\item Run the {\sc UnrankDyck} procedure in Algorithm~\ref{UnrankDyck} to construct the Dyck word $w$ of semilength $n$ and rank $p$.
\item Using the considerations above, construct the set $\{(m_i,\pi(m_i))\}_{i=1}^k$ of left-to-right maxima (values, positions) of the permutation $\Psi^{-1}(w)$.
\item Finally, construct the permutation $\unrank_{\Av_n(321)}(p)$.
\end{itemize}
\end{proposition}

\section{Trivial symmetries}

Here, based on the previous results and the trivial symmetries between permutation classes, we derive ranking and unranking algorithms for the remaining length-3 avoiding patterns: 123, 132, 213, and 312.

Let $\mathfrak c$ be the complement operation on $\mathfrak{S}_n$, that is the involution  defined as:
$\sigma=\mathfrak c(\pi)$ if  $\sigma(i)=n+1-\pi(i)$ for $1\leq i\leq n$.
\noindent
Clearly, the complement operation $\mathfrak c$ induces a bijection between 
$\Av_n(231)$ and $\Av_n(213)$, 
and between $\Av_n(321)$ and $\Av_n(123)$. Moreover, if $\pi<\rho$ in lexicographic order, then $\mathfrak c(\rho)<\mathfrak c(\pi)$ in lexicographic order.
With the considerations above we have:

\begin{proposition}
Let $\tau$ be one of the length-3 patterns $213$ or $123$ (and so, $\mathfrak c(\tau)$ is $231$ or $321$). Then, for $n\geq 0$ and 
\begin{itemize}
\item for $\pi\in \Av_n(\tau)$,
$\rank_{\Av_n(\tau)}(\pi) = c_n-\rank_{\Av_n(\mathfrak c(\tau))} (\mathfrak c(\pi))-1$, and
\item for an integer $p$, $0\leq p<c_n$,
$\unrank_{\Av_n(\tau)}(p) = \unrank_{\Av_n(\mathfrak c(\tau))}(c_n-p-1)$. 
\end{itemize}
\end{proposition}

Let now $\mathfrak r$ be the reverse operation on $\mathfrak{S}_n$, that is the involution on $\mathfrak{S}_n$ defined as:
$\sigma=\mathfrak r(\pi)$ if  $\sigma(i)=\pi(n+1-i)$ for $1\leq i\leq n$.
The reverse operation $\mathfrak r$ induces a bijection between 
$\Av_n(231)$ and $\Av_n(132)$, 
and between $\Av_n(213)$ and $\Av_n(312)$. Moreover, if $\pi<\rho$ in lexicographic order, then $\mathfrak r(\pi)<\mathfrak r(\rho)$ in colexicographic order. Denoting $\rank^*$ (resp. $\unrank^*$) the ranking (resp. unranking) function in colexicographic order
we have:

\begin{proposition}
Let $\tau$ be one of the length-3 patterns $132$ or $312$ (and so, $\mathfrak r(\tau)$ is $231$ or $213$). Then, for $n\geq 0$ and
\begin{itemize}
\item for $\pi\in \Av_n(\tau)$, $\rank^*_{\,\Av(\tau)}(\pi)=\rank_{\,\Av(\mathfrak r(\tau))}(\mathfrak r(\pi))$,
\item for an integer $p$, $0\leq p<c_n$, $\unrank^*_{\,\Av(\tau)}(p)=\mathfrak r(\unrank_{\,\Av(\mathfrak r(\tau))}(p))$.
\end{itemize}
\end{proposition}
\medskip

\noindent
{\bf Open problems and future directions:}

Even when the counting sequences for permutations avoiding multiple patterns of length three are trivial (e.g., \(2^{n-1}\), Fibonacci numbers), ranking and unranking methods in lexicographic order still seem less straightforward. It would also be interesting to consider the avoidance of patterns of size greater than three. Finally, the efficient implementation of these methods remains an open problem.




\nocite{*}
\bibliographystyle{eptcs}
\bibliography{biblio}
\end{document}